# Sunspot observations by Charles Malapert during the period 1618–1626: a key dataset to understand solar activity before the Maunder minimum

V.M.S. Carrasco[1,2], M.C. Gallego[1,2], J. Villalba Álvarez[4], J.M. Vaquero[2,3]

[1] Departamento de Física, Universidad de Extremadura, 06071 Badajoz, Spain [e-mail: vmscarrasco@unex.es]

[2] Instituto Universitario de Investigación del Agua, Cambio Climático y Sostenibilidad (IACYS), Universidad de Extremadura, 06006 Badajoz, Spain

[3] Departamento de Física, Universidad de Extremadura, 06800 Mérida, Spain

[4] Departamento de Ciencias de la Antigüedad, Universidad de Extremadura, 10071 Cáceres, Spain

**Abstract:** A revision is presented of the sunspot observations made by Charles Malapert from 1618 to 1626, studying several documentary sources that include those observations. The revised accounting of the group numbers recorded by Malapert for that period shows new information unavailable in the current sunspot group database. The average solar activity level calculated from these revised records of Malapert is by almost one third greater than that calculated from his records included in the current group database. Comparison of the sunspot observations made by Malapert and by other astronomers of that time with regard to the number of recorded groups and sunspot positions on the solar disk shows good agreement. Malapert reported that he only recorded one sunspot group in each sunspot drawing presented in *Austriaca Sidera Heliocyclia* (the documentary source which includes most of the sunspot records made by Malapert), although he sometimes observed several groups. Therefore, the sunspot counts obtained in this present work on Malapert's sunspot observations represents the lower limit of the solar activity level corresponding to those records.

## 1. Introduction

The sunspot number is the index most used to describe the variability of the long-term solar activity (Vaquero, 2007; Usoskin, 2017). This index is calculated from the sunspot records available for approximately the last 400 years, corresponding to the telescopic era (Hoyt & Schatten, 1998; Clette et al., 2014). Several recent studies have detected problems in both some historical sunspot records and the methods used to reconstruct the sunspot number (Clette et al., 2014; Carrasco et al., 2015). Thus, a new revised collection of the number of sunspot groups (Vaquero et al., 2016) and several new sunspot series (Clette & Lefèvre, 2016; Lockwood et al., 2016; Svalgaard & Schatten, 2016; Usoskin et al., 2016; Chatzistergos et al., 2017; Willamo et al., 2017) have been published in order to resolve those problems. Recently, Muñoz-Jaramillo & Vaquero (2018) showed how hard it is to connect the modern sunspot observations with the historical data due to the low temporal coverage in the sunspot records belonging to the first two centuries of the telescopic era. Nowadays, the sunspot number community is making an effort to improve these sunspot group databases, and to reach a widely accepted version of the sunspot number index.

Recently, Carrasco et al. (2019) demonstrated that Charles Malapert sometimes recorded several sunspot groups as only one representative group. As a continuation to that work, we here analyse the sunspot observations made by Malapert during the first quarter of the 17th century (Malapert, 1620, 1633; Scheiner, 1630). The main documentary source in which most of the sunspot observations made by Malapert are included, Austriaca Sidera Heliocyclia (Malapert, 1633), was published three years after Rosa Ursina (Scheiner, 1630), regarded as one of the most thorough surveys of sunspots of that time (Vaquero & Vázquez, 2009). Moreover, Scheiner (1630) provided sunspot observations made by Malapert in 1624 and 1625 which are not included in the recent sunspot group databases (Hoyt & Schatten, 1998; Vaquero et al., 2016). Malapert was, together with Christoph Scheiner (Daxecker, 2004, 2005), the most active sunspot observer of his time (Vaquero et al., 2016). Sunspot observations made by Malapert are of especial interest due to the low temporal coverage of the sunspot records in that period. Moreover, according to Vaquero et al. (2016), Malapert was the only observer on roughly 60% of the days when he recorded observations. Figure 1 shows the number of sunspot groups recorded by all observers during the period 1618–1626 obtained from Vaquero et al. (2016). One can see in the figure that, although sunspot observations are available for each year of that period, the 11-year solar cycle shape is unclear because the daily number of sunspot groups recorded from 1618 until the end of 1624 was always equal to one. This period is very important to constrain models of switching the solar dynamo between regular cyclic and grand-minimum modes (Vaquero et al., 2011).

We note that the sunspot observations made by Malapert were previously analysed by Hoyt & Schatten (1998) and Neuhäuser & Neuhäuser (2016), and Vaquero et al. (2016) contains the same sunspot observations recorded by Malapert as Hoyt & Schatten (1998). However, in this work, we carried out a revision of these important historical data after translating the original Latin, including sunspot records made by Malapert in 1624 and the first third of 1625 previously not analysed.

This work is part of the effort to improve the current issues of sunspot group databases. Its objective is to perform an analysis of the sunspot records made by Malapert during the period 1618–1626. The original texts, together with translations, and the new group number count are available on the Historical Archive of Sunspot Observations website (HASO, http://haso.unex.es). Section 2 is devoted to some general information about Charles Malapert. We present the observation method and data recorded by Malapert in Sections 3 and 4, respectively. We discuss and analyse the results in Section 5, and present the main conclusions in Section 6.

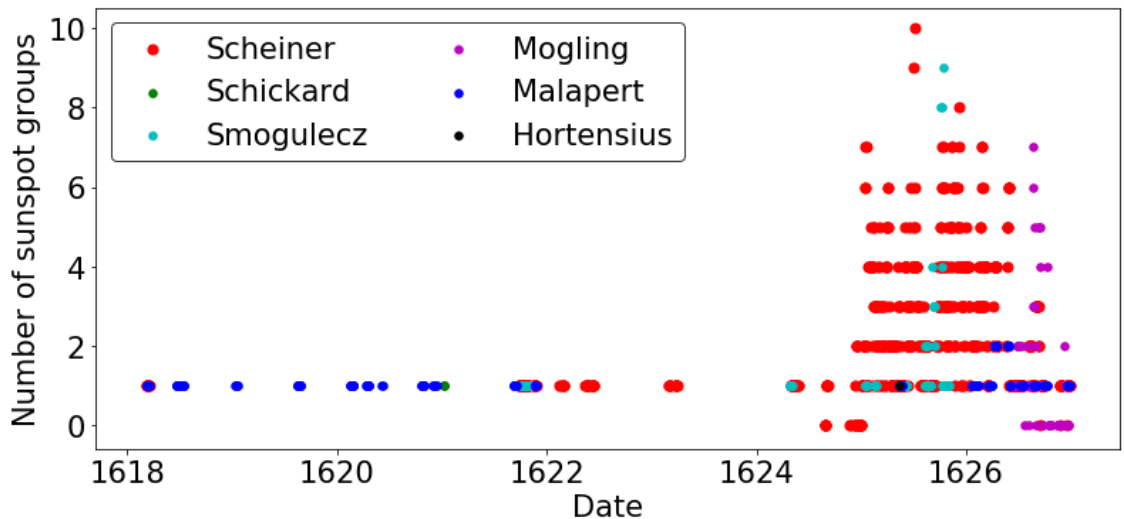

Figure 1. Daily number of sunspot groups recorded by all the observers available in the group database of Vaquero et al. (2016) during the period 1618–1626.

## 2. Charles Malapert

Charles Malapert was born in Mons (Southern Netherlands, under the Roman Catholic King of Spain, nowadays Belgium) in 1581 (Quignon, 1930; Bloemendal & Norland, 2013). He entered the Jesuit order in 1600, and started to teach Mathematics and Philosophy in Mons. After teaching in Lorraine and Kalisz, he was sent to Douai in 1617 to be a professor at the University of Douai. At the end of the 1620's, Malapert became rector of Arras (France). He died in Vitoria (Spain) in 1630 when he was traveling to Madrid to teach Mathematics in a newly created chair at the Imperial College of the Society of Jesus of Madrid (Birkenmajer, 1967).

Malapert published works of diverse genres, from poetry to scientific books (Quignon, 1930; Mertz et al., 1990). In the scientific field, Malapert stood out for his books on mathematics and astronomy. He carried out studies about the Moon (Malapert has a lunar crater named after him), comets, and sunspots. Malapert performed his sunspot observations together with his assistant Sylvius Polonus at Douai (50°22'15" N 3°04'45" E) in the region of Flandes (France) during the period 1618–1626 (Birkenmajer, 1967). Most of their sunspot records were published in Austriaca sidera heliocyclia (Malapert, 1633). The sunspot observations made by Malapert in March 1618 were also published in Oratio habita Duaci dum lectionem mathematicam auspicaretur (Malapert, 1620). Lastly, the observations performed in 1624 and the first third of 1625 were published by Scheiner (1630). The book written by Malapert (1633), published three years after Malapert's death, is devoted completely to the study and discussion of the sunspot observations. We would note that Malapert (1633) is one of the first documentary sources available that includes extensive information about sunspot observations and a discussion concerning their nature.

Just after the telescope began to be used as an astronomical instrument, an important discussion took place among astronomers of that time about the nature of sunspots. We would like to highlight a comment included in Malapert (1620) in this discussion as to whether or not sunspots are actually located on the solar disk: [Original text] "… Ea sane quaecunque demum sint corpora satis constat longe supra Lunam suos gyros ducere, quae toto diei cursu sub Sole perseverent totique Europae eodem temporis articulo sub eadem Solis parte visantur; quod ego literis ab ultima Polonia et aliunde acceptis certum reddere possum atque testatum. Quin et illud mihi certe persuasum est, circulo circa Solem maculas hasce converti, cum tardiores et confertae magis appareant circa oras extremas, quae deinde medium sub Solem explicant sese feruntque celerius." [English translation] "… Whatever these bodies are finally, it is quite evident that they go around far above the Moon, that they remain under the Sun during the entire course of the day, and that they are observed throughout Europe at the same moment and under the same part of the Sun. This I can give assurance for as being true and proven through letters that I have received from distant Poland and from other places. Moreover, I am fully convinced that these spots go around in a circle close to the Sun, since they appear to be slower and more compact at the outer edges, and then grow in extent and move faster in the middle of the Sun." Therefore, it seems that Malapert defended the idea that sunspots were celestial bodies orbiting close to the Sun.

## 3. Observation Methods

Malapert (1633) stated that a lot of celestial objects unknown in the past can be observed through the telescope, including sunspots. He explained different methods to observe sunspots (see pp. 21-28 of Malapert, 1633). He pointed out that one or two green or dark lenses must be installed in the telescope in order for the eyes not to be damaged by the brightness of the Sun. He also mentioned that the Sun can be observed directly through the telescope without those dark lenses, for example, if there is a dense fog (a practice that is of course inadvisable due to the potential problems that it can cause to the eyes). Moreover, Malapert noted that sunspots can be observed without a telescope by passing the solar light through a pin-hole into a dark place and projected onto paper. Nonetheless, Malapert indicated that the best method to observe sunspots is through the telescope since only the largest sunspots can be observed without a telescope.

Unfortunately, Malapert (1620, 1633) did not provide information about the optical characteristics of the telescopes used for his sunspot observations. The instrument employed by Malapert to observe sunspots is shown in *Austriaca Sidera* (Figure 2). He considers that the method used to observe with that instrument is, according to his experience with the manufacture and use of telescopes, not simple. The method consists in projecting the Sun's image onto a sheet of paper, taking into account the following points (see Figure 2): (i) a column A must be leveled such that the side CD constitutes the angle of elevation of the equator with the line DE, parallel to the horizon (Malapert pointed out that the angle CDE for his location is 39.5° [colatitude]), and a circle's axis H must be placed in the middle of that surface; (ii) the upper plane KNOP must have an angle of maximum inclination around 23.5° (the obliquity of the ecliptic) with respect to the lower plane KLM; (iii) the telescope is inserted into the boards V and T, mounted on a long slat IT at right angles, projecting the Sun onto I which is located at the appropriate distance to observe sunspots; (iv) the surface DEFG must be in the meridian plane, and the angle CDE rises in the opposite direction at noon; (v) no light source must impinge on panel I, so that the observer should cover their head with a cloak in such a way that the panel I is also covered; and (vi) in order to have the same size disk throughout the year, the distance between the telescope and panel I should be reduced in summer and extended in winter to compensate for the seasonal difference in the size of the solar disk.

Malapert (1633, pp. 25–28) pointed out that if sunspots are observed on successive days then it can be seen that their trajectory is parallel to the ecliptic line SX (see Figure 2). Furthermore, Malapert (1633, pp. 150-151) noticed the nonlinear path followed by some sunspots crossing the solar disk, indicating that "… sunspots sometimes deviate from the initial trajectory due to movements of the epicycles." However, he did not identify sunspots as describing those paths in months around the equinoxes. Malapert added "… sunspots occasionally seem to move faster or slower than normal due to mistakes in the dates or the identification of the sunspot studied." For example, Malapert (1633, pp. 150-151) explained that if a spot A is observed (next to the solar limb) for a given day and another one B (closer to the centre of the solar disk) cannot be seen because it is too dissipated, then it may happen that, on the following day, the spot B becomes visible close to the centre of the solar disk while the spot A cannot be seen that day, so that someone might think that spot A has advanced more than normal from a point near the solar limb to a point near the centre in just one day. Malapert also affirmed that he never saw a sunspot that moved backwards or even remained motionless.

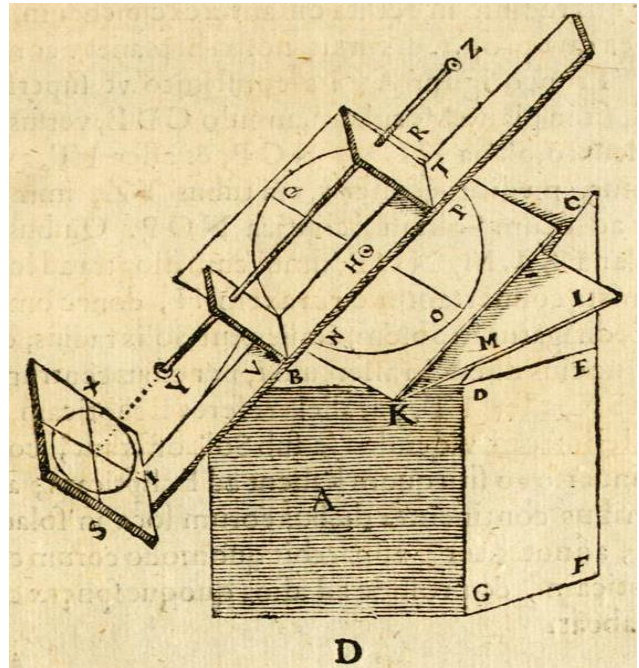

Figure 2. Instrument used by Charles Malapert to observe sunspots [Source: Malapert, 1633, p. 25].

## 4. Data

Malapert carried out his sunspot observations during the period 1618–1626. Together with Christoph Scheiner, he was the most important observer of that time in terms of the number of sunspot records (Vaquero et al., 2016). Malapert published 28 sunspot drawings in two books (Malapert, 1620; 1633). The only sunspot drawing about the observations made in March 1618 that was published in Malapert (1620) was also included in Malapert (1633). Those sunspot observations corresponding to March 1618 were also published by Scheiner (1630, p. 229). Moreover, another three sunspot drawings made by Malapert in 1624 and two in 1625 were published by Scheiner (1630) corresponding to the periods: (i) 30 April to 10 May 1624 (Scheiner, 1630, p. 229), (ii) 28 July to 12 August 1624 (Scheiner, 1630, p. 257), (iii) 31 August to 5 September 1624 (Scheiner, 1630, p. 281), (iv) 26 January to 7 February 1625 (Scheiner, 1630, p. 181), and (v) 24 March to 2 April 1625 (Scheiner, 1630, p. 195). We would note that Malapert published one sunspot drawing (Malapert, 1633, p. 70) that included sunspot observations made in Ingolstadt (Germany), as well as another two sunspot drawings (Malapert, 1633, pp. 70 and 75) made by Simon Perovius in Kalisz (Poland). Malapert did not name the astronomer who performed the sunspot observations at Ingolstadt and Kalisz in March 1618. Instead, he indicated that Perovius was responsible for the sunspot records made in Kalisz in July 1618. According to Quignon (1930) and Vaquero & Vázquez (2009), Johann Cysat carried out the Ingolstadt sunspot observations that were published by Malapert (1633).

An important difference between the three documentary sources consulted for this work is that sometimes several sunspot groups were drawn for the same day by Malapert (1620) and Scheiner (1630). Malapert (1633) drew only one sunspot group in each drawing, describing its trajectory across the solar disk (Figure 3). Moreover, one can see that he sometimes depicted several individual sunspots in the same group. Malapert (1633) also added a textual description of each group to accompany the sunspot drawings. The information included in the sunspot drawings is: (i) the horizontal line, delimited by the letters A and B, depicts the ecliptic plane; (ii) letters C and D label the northern and southern solar hemispheres; (iii) the black dots represent the sunspot groups (or only one representative sunspot group if several groups were observed on the same day); and (iv) the upper and lower series of numbers indicate the day and hour of the observation, respectively (the meridian altitude is sometimes given). We note that if discontinuous days appear in the drawings then it is because Malapert could not observe on the missing days, so that they are not necessarily spotless days. Furthermore, it can be seen in each chapter of Malapert (1633) that the sunspot drawings follow a monthly order, starting with observations made in January and finishing in December, independently of the year.

Sunspot drawings and textual records made by Malapert (1620; 1633) and Scheiner (1630) in accordance with Malapert's observations are carefully analysed in this present work. We have translated the original Latin texts and prepared a new accounting of the group numbers recorded by Malapert. The original Latin texts and their translations are available on the website of the Historical Archive of Sunspot Observations (HASO, http://haso.unex.es). Table 1 presents the annual number of sunspot records made by Malapert and the average of the group number for the period 1618–1626 according to this work and to the current group database (Vaquero et al., 2016). We note: (i) this new revision contains important differences with the group databases, (ii) Malapert did not record sunspots in 1622 and 1623, and (iii) the statistics corresponding to Malapert`s records given in Table 1 from Vaquero et al. (2016) also include observations made by Wely (at Coimbra), Cysat (at Ingolstadt), and Perovius (at Kalisz). These facts will be discussed in the next section.

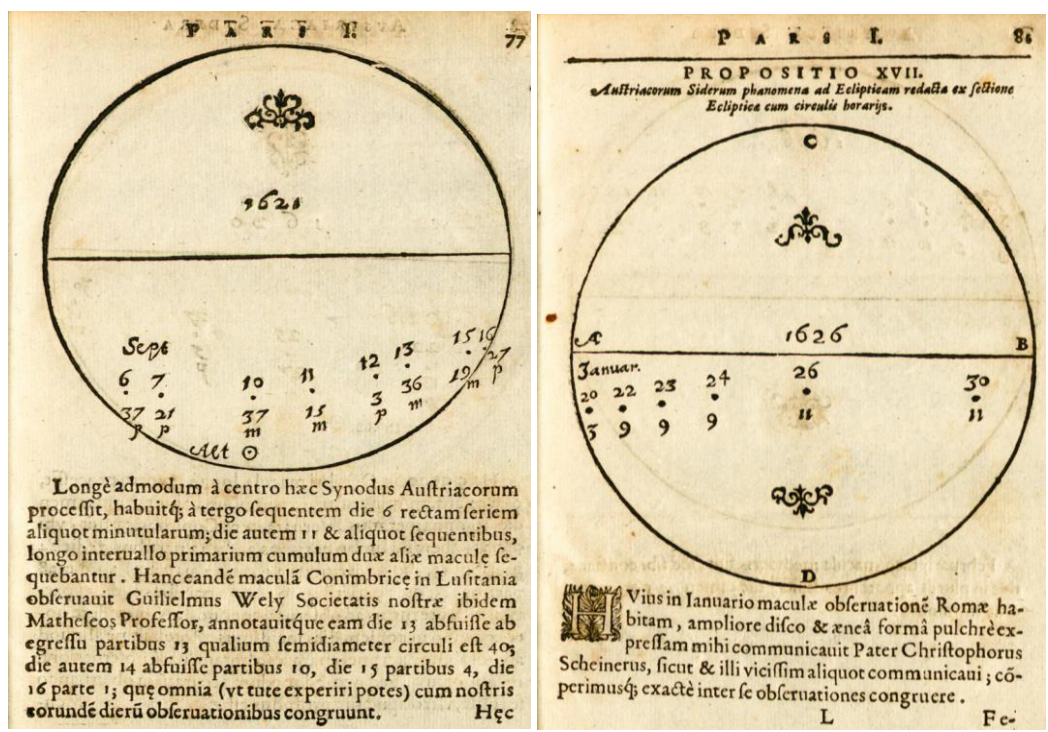

Figure 3. Two example pages with sunspot drawings made by Malapert [Source: Malapert, 1633, pp. 77 and 81].

Table 1. Number of annual sunspot observations and average of the group number recorded by Malapert for the period 1618–1626 according to this work and to the current group database (Vaquero et al., 2016).

| YEAR | THIS WORK | | GROUP DATABASE | |
| --- | --- | --- | --- | --- |
| | NUMBER OF RECORDS | AVERAGE OF THE GROUP NUMBER | NUMBER OF RECORDS | AVERAGE OF THE GROUP NUMBER |
| 1618 | 23 | 1.3 | 13 | 1 |
| 1619 | 20 | 1.2 | 20 | 1 |
| 1620 | 28 | 1 | 37 | 1 |
| 1621 | 13 | 1.9 | 13 | 1 |
| 1624 | 35 | 1.1 | 0 | - |

| 1625 | 41 | 2.1 | 12 | 1 |
| 1626 | 91 | 1.5 | 82 | 1.2 |

## 5. Results and Discussion

### 5.1. Malapert's Data

Malapert observed on 251 different days spread over 7 years. It is not a great number of observations but it is very important because, as mentioned above, Malapert was the only observer in the roughly 60% of his observation days according to the current group database. In general, Malapert only reported days with spots on the Sun. The only cases when Malapert explicitly reported spotless days were from 27 to 30 August 1624 and 7 September 1624 (Scheiner, 1630, p. 280). This information is only available from the textual report. The year with most records made by Malapert was 1626 (91 records). The minimum and maximum of this solar cycle lie within 1620 and 1625, respectively, according to the average of the group number. We can see that values of the yearly average of the group number do not present gradual changes, as expected in a standard 11-year solar cycle with a rise and decline phase. The low annual number of sunspot records and the difficulty in obtaining the real number of sunspot groups observed by Malapert (from drawings and, mainly, texts) could be factors that contribute to the changes in the yearly average of the group number not being gradual.

We found a possible erratum in the textual report corresponding to the sunspot drawing of November 1621. In that drawing (Malapert, 1633, p. 79), Malapert reported that several sunspots were observed from 20 November, but only one remained on 30 November. However, the last day recorded for that sunspot in the drawing is on 25 November. Thus, we considered that the information corresponding to 30 November actually corresponds to 25 November, and decided to include no information about 30 November 1621 in the new group number accounting. We note that, to obtain the total number of groups recorded by Malapert from 30 August to 3 September 1626, we had to sum the number of groups recorded in two different drawings (Malapert, 1633, pp. 88-89) since those drawings include different sunspots recorded on those dates. It must also be taken into account that Malapert reported that the sunspot group recorded in the drawing from 23 August to 2 September 1626 was also seen on the solar limb on 3 September. This fact is not reflected in the drawing, only in the text (Malapert, 1633, p.

88). The appearance of sunspots on 6 and 7 June 1620 is also reported only in the text (Malapert, 1633, pp. 150-151). Furthermore, Malapert also indicated that he only observed that sunspot on those two days, and, unlike Neuhäuser & Neuhäuser (2016), we decided not to consider the day before and after as spotless days because that information is not explicitly reported by Malapert in the text.

Malapert mentioned the appearance of faculae in February 1620 and March 1625. He indicated that faculae usually appeared behind sunspot groups on the solar limb (Malapert, 1633, p. 68): [Original text] "Sub ingressum die 17 affusae erant a tergo faculae, ut et circa alios cumulos ad ingressum et egressum saepius videri solent." [English translation] "At its [a sunspot group's] entrance, on the 17th, faculae appeared behind as often appear around other cumuli at the entrance and exit [from the solar disk]."

## 5.2. The sunspot count recorded by Malapert

The main documentary source (Malapert, 1633) about the sunspot observations made by Malapert provides important information about the number of groups recorded by this astronomer. We concluded from the text that Malapert always represented just one sunspot group in his sunspot drawings although he sometimes observed several sunspot groups: [Original text] "neque enim, ut ante iam monui, minutias omnes pingere necessarium duxi, quin immo cum plures et inter se distantes simul apparuerunt saepe cumuli, unici tamen cursum hic repraesentavi." [English translation] "And, as I already mentioned, I did not consider it necessary [in the sunspot drawings] to draw all the particles; in addition, when many times several cumuli appear and moreover far from each other, I have represented here the trajectory for only one." This key point to understanding Malapert's records was not taken into account in the sunspot group database (Vaquero et al., 2016). Instead, we would note that Malapert recorded more than one individual sunspot in the same group in several cases (Carrasco et al., 2019).

Descriptions made by Malapert about his observations are sometimes unclear. Thus, it is no trivial matter to obtain the number of groups from those textual reports. An example of this fact is the report made by Malapert (1633, p. 78) corresponding to the sunspot observation in October 1620: [Original text] "Haec quoque in Octobri macula longe a centro processit, et sequentem minutularum cumulum habuit, qui tamen die 30 disparuerat." [English text] "This spot of the month of October also moved away from the centre, and had behind a cumulus of tiny spots, which however had disappeared on the 30th." In this case, we consider that Malapert observed only one sunspot group, although, from this comment, it is not very clear if there were one or more sunspot groups on the Sun. Thus, in order to count the number of sunspot groups, we apply the following criteria: (i) if Malapert pointed out that he observed one group but the group broke into two or more "cumuli" on a given day, we consider only one group unless Malapert reported a certain distance between them, and (ii) if Malapert reported sunspots above or below other sunspots then they are regarded as different groups. An example where we apply the first criterion corresponds to the sunspots recorded in September 1621 (Malapert, 1633, p. 77): [Original text] "… die autem 11 et aliquot sequentibus, longo intervallo primarium cumulum duae aliae maculae sequebantur…" [English translation] "… however, on the 11th and successive days, this first cumulus was followed by two other spots at a large distance…". An example of the second criterion can be found in the sunspot observations made in April 1626 (Malapert, 1633, p. 84): [Original text] "… Infra hunc cumulum duae exiguae maculae cursum huic parallelum tenebant, ut et alias fieri assolet." [English translation] "… Under this cumulus, two small spots followed its same path, in parallel, as usually occurs many other times." We would note that we also applied the second criterion to the sunspot observed in November 1621 (Malapert, 1633, p. 79), but the textual report is not clear as to whether one or more sunspot groups were observed by Malapert in that period (Malapert, 1633, p. 79): [Original text] "Mense Novembri hic cumulus die 20 quatuor habuit recta serie prope cohaerentes maculas exiguas, et quintam tertiae suppositam…". [English translation] "In the month of November, this cumulus presented on the 20th four small spots joined in a straight line, and a fifth under the third…".

### 5.3. Comparison with the group databases

We recovered sunspot observations performed by Malapert not included in the sunspot group database (Vaquero et al., 2016). The number of Malapert's records analysed in this work is therefore greater than that included in the group databases (Table 1). In particular, in this work we include 88 records made by Malapert which were not taken into account in the group database. These records correspond to the sunspot drawings of April-May, July-August, and August-September 1624, January-February, March-April, October, and November 1625, and June-July 1626, as well as the spotless day of 7 September 1624. We also corrected dates for the sunspot drawing of September 1626 (Malapert, 1633, p. 89) because the current group database considers the sunspot observations recorded in that drawing to have been made in October 1626. Furthermore, from the textual report in Malapert (1633), we realized that nine sunspot observations made by Wely were erroneously assigned to Malapert in the group databases, as shown by Neuhäuser & Neuhäuser (2016). Instead, the sunspot observations by Cysat at Ingolstadt and by Perovius at Kalisz reported by Malapert (1633) were not included in the database. The group database may also have erroneously assigned to Scheiner the sunspot observations made by Malapert in March 1618 and August-September 1624. Figure 4 shows the daily number of sunspot groups recorded by Malapert during the period 1618–1626 according to the group database (top panel) and this work (bottom panel). We note that the bottom panel of the figure includes the new and corrected information about records made by Malapert, Cysat, Perovius, and Wely.

Values of the yearly average of the group numbers obtained in this new accounting are greater than those obtained from records included in the current group database, except the value for 1620 where both averages are equal. This fact is because, according to the group database, the daily number of groups recorded by Malapert is generally one. The only case with a group number different to that value corresponds to the sunspot observations made in April and May 1626, where the number of sunspot groups recorded was 2. In general, we also obtained higher daily accountings of the number of groups, up to a maximum value of five groups. Furthermore, the average solar activity level obtained in this work (1.44) calculated from the average of the group number recorded by Malapert for the period 1618–1626 is greater than those obtained from the current group database (1.09).

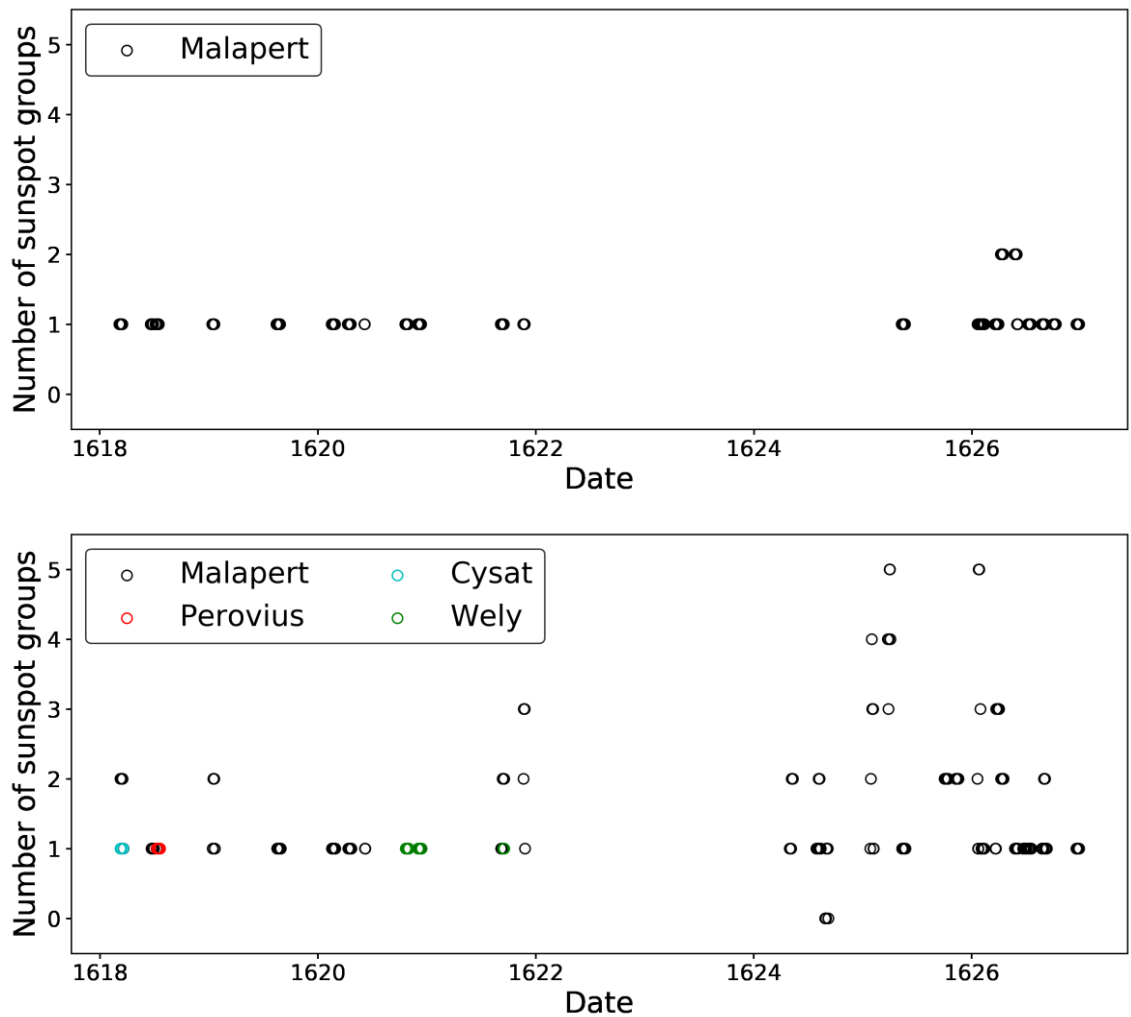

Figure 4. Daily number of sunspot groups recorded by Malapert (black color) during the period 1618–1626 according to the current group database (top panel) and this work (bottom panel). We would note that some observations assigned to Malapert in the group database were actually carried out by Cysat (cyan color), Perovius (red color), and Wely (green color).

### 5.4. Comparing observations

First, we compare the observations made by Malapert for the same period (March 1618) published in different documentary sources (Figure 5). We found an evident difference between records published in Malapert (1620) and Malapert (1633): the former recorded two different sunspot groups for that period while the latter only one group. Furthermore, these sunspot observations made by Malapert in March 1618 were also published by Scheiner (1630, p. 229). That sunspot drawing agrees with Malapert (1620) since the groups A and B are also shown. Scheiner wrote: [Original text] "Tabula

I. Complectitur cursus macularum a & b anno 1618 peractos, quorum mihi observationes omnes praeter diem 15 hora 12 peractas, misit P. Carolus Malapersius; factae sunt ab octavo ad 18 Martii, die 15 easdem observavit hora 2 pomeridiana. Edidit easdem ipsemet olim, idem Pater in suo solemni ad doctrinam mathematicam initio, quod Euclidi in fine adiunxit, sed forma multo contractiore. Hanc autem cuius apographum ego tibi do, misit ad me, sua manu conscriptam, quam proinde studiose servo. Maculas porro a ea figura et magnitudine tibi propono, qua ab illo accepi. Etenim nec addere nec demere quidquam libebat alienis; meas porro huius anni millesimi sexcentesimi decimi octavi observationes in Germania peractas, hic ad manum non habeo, certus nihilominus sum neque has ab illis neque illas ab hisce loco et situ discrepare." [English translation] "Table 1. Includes trajectories followed by sunspot *a* and *b* in 1618. Fr. Charles Malapert sent me these observations and they were obtained at 12:00 noon, except on the 15th. He carried them out from the 8th to the 18th of March, those of the 15th were observed at 2:00 p.m. That same Father published them some time ago, at the solemn beginning of his Mathematics lessons, which he added at the end of his Euclid [Malapert, 1620], although in a much more summarized way. This version of which I give you a copy he sent me written in his own hand, and of course I keep it with care. Furthermore, I reproduce spots a with the same shape and size with which he sent them to me. I certainly did not want to add or remove anything from what was not mine. Regarding the observations that I made in Germany in that same year 1618, I do not have them at hand, but I am sure that neither do these differ from those nor those from these as to their place and position." According to this comment, we can confirm that Malapert observed two sunspot groups (*a* and *b*) in March 1618, although he only reported one group in Malapert (1633).

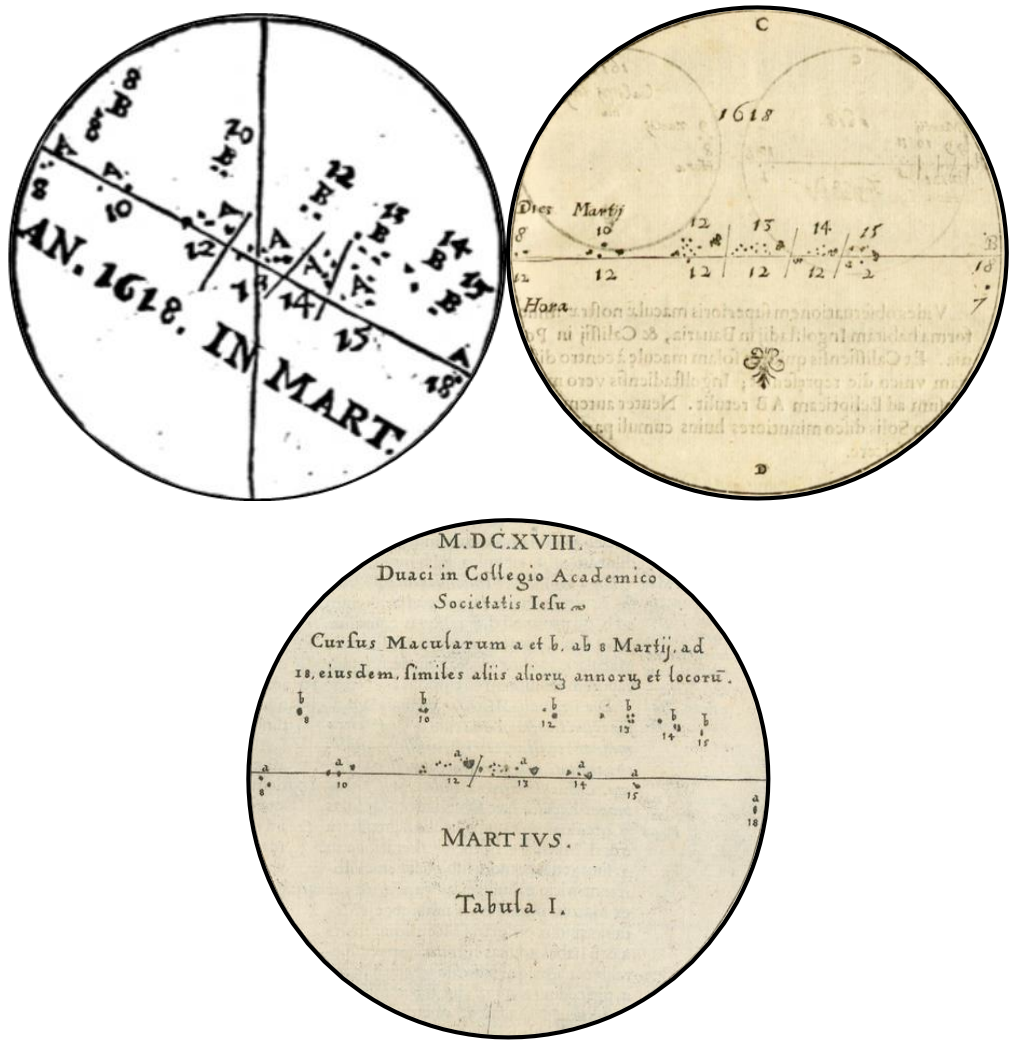

Figure 5. Sunspot drawings recorded by Malapert in March 1618 included in: i) upper-left panel – Malapert (1620, p. 22), ii) upper-right panel – Malapert (1633, p. 69), and iii) bottom panel – Scheiner (1630, p. 229).

We have also compared observations made by Malapert and other observers for similar dates. First, we compared the sunspot observations made by Malapert and Christoph Scheiner (Scheiner, 1630) in January 1626. Malapert (1633, p. 81) indicated in a textual report (Carrasco et al., 2019): [Original text] "Huius in Ianuario maculae observationem Romae habitam, ampliore disco et aenea forma pulchre expressam mihi communicavit Pater Christophorus Scheinerus, sicut et illi vicissim aliquot communicavi; comperimusque exacte inter se observationes congruere." [English translation] "The observation of this spot in January, which took place in Rome, on a larger and beautifully engraved bronze disk, was shared with me by Father Christoph Scheiner, with whom I have also shared a certain number, and we were able to verify that the observations of both of us coincided exactly." Despite this comment, only one sunspot group was recorded by Malapert (1633) for each observation day from 20 to 30 January 1626, while Scheiner recorded between 3 and 5 groups on the days that both astronomers were observing. We note that, in order to obtain the total number of groups recorded by Scheiner for each day of that period, we summed the number of sunspot groups included in two different drawings (Scheiner, 1630, pp. 295 and 297). According to this agreement between sunspot records made by Malapert and Scheiner, in our new group accounting of Malapert's records, we assigned to Malapert the same group number recorded by Scheiner for those days when both astronomers were observing, since Scheiner recorded the total number of groups observed. We can also compare the observations made by Malapert and Scheiner from sunspot drawings (Scheiner, 1630, pp. 193 and 195). Figure 6 shows the sunspot groups recorded by Malapert and Scheiner for the same observation period, from 24 March to 2 April 1625. One can see the similarity of these observations in both the position and the number of groups recorded. The only difference between the observations is on 31 March, when Malapert recorded two groups and Scheiner five groups. However, it can be seen that groups d, e, and f were omitted in the drawing corresponding to Malapert's observations for that date, but, instead, they were recorded on the day before and the day after. It thus seems to be a transcription error. The resulting accounting of groups recorded by Malapert and Scheiner from 24 March to 2 April is then exactly the same.

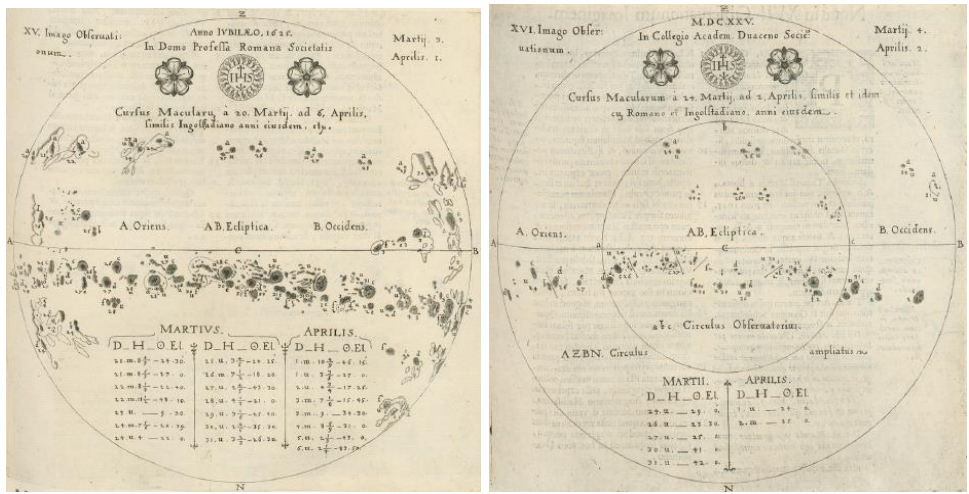

Figure 6. Sunspot drawings made by Scheiner (left panel) and Malapert (right panel) including records for the same observation period, from 24 March to 2 April 1625 [Source: Scheiner, 1630, pp. 193 and 195].

Scheiner (1630) also included a comparison of the sunspot observations made by himself at Rome, Malapert at Douai, and Schönberger at Ingolstadt in January-February 1625 (Figure 7, top panel). These three observations present some differences. With Scheiner's records as referents, one can see that Malapert did not record the group *b* observed by Scheiner on the solar limb on 26 January, groups c and d on 28 January, and groups a (close to the solar limb) and *c* on 6 February. Also, Schönberger did not record groups *c* and *d* on 27 January, or group *a* (close to the solar limb) on 6 February. The only difference in the three coincident observation days (31 January, and 5 and 6 February) of the observations by Malapert and Schönberger corresponds to 6 February when Schönberger registered group *c* but Malapert did not. Furthermore, group *a* observed by Scheiner (1630, p. 183) from 1 to 12 February is not present in the sunspot drawings made by Malapert and Schönberger. Given the size recorded by Scheiner for that group, it should have been observed by both Malapert and Schönberger. This would indicate that, in order for these observations to be compared, not all the groups observed in that period by Malapert and Schönberger were included in the sunspot drawings, but only groups *a*, *b*, *c*, and *d*. Indeed, Scheiner makes a statement (1630, p. 194) confirming this: [Original text] "… Secundo, ob vitandam confusionem e locorum penuria, non omnes observationum dies aut maculas adduxi, sed opportuniores, quibus mirifica tantorum locorum consonantia in tam dissita re constare…". [English translation] "… Second, in order to avoid confusion caused by the lack of space, I have not presented every observing day or sunspot, but only those sunspots most opportune to let the amazing correspondence of so many positions in such a scattered matter be known…".

Scheiner also published a sunspot group crossing the solar disk from 30 April to 10 May 1624 observed by Schönberger and Malapert (Scheiner, 1630, pp. 217 and 229). Both observers recorded just one individual sunspot in their sunspot drawings for each observation day, with a very similar trajectory across the solar disk for that group (Figure 7, bottom panel). However, Scheiner (1630, p. 228) reported: [Original text] "Unum ferme oblitus essem, nimirum die 6 Maii ante centrum, eidem observatori

Duaceno repente novam comparuisse Maculam, antea ut ipse scribit non visam, quam deinde ad finem usque continuat. Cum autem Ingolstadiano dies sextus non favisset, vidit eandem valde magnam non procul post centrum versus occasum, quam ad finem usque per observationes diurnas est pariter prosecutus. Cursum illius, sicut multarum aliarum non appono, sed moneo haec, ut advertas non soli mihi sed aliis quoque in Sole medio maculas novas nasci et antiquas interire." [English translation] "… I almost forgot one thing, it is certain that the same observer of Douai suddenly on the 6th of May saw a new spot, as he writes not seen before, in the centre [of the Sun], and then it continued until its end. For the observer of Ingolstadt, since the 6th was not suitable, he saw that same large spot not far in the setting direction from the centre, and also followed it in daily observations until its end. I do not give its course, as neither do I for many other spots, but I advise you of this so that you note that not only I but also others think that new spots appear and old ones disappear in the middle of the Sun." Therefore, although not shown in the drawings, Malapert and Schönberger observed two groups from 6 May. We thus consider two sunspot groups from 6 to 10 May in this new accounting in accordance with Malapert's observations. Furthermore, we would note that it is not possible to compare the other two sunspot drawings made by Malapert and published by Scheiner (1630) with drawings recorded by other observers because they are the only sunspot observations available for those observation days.

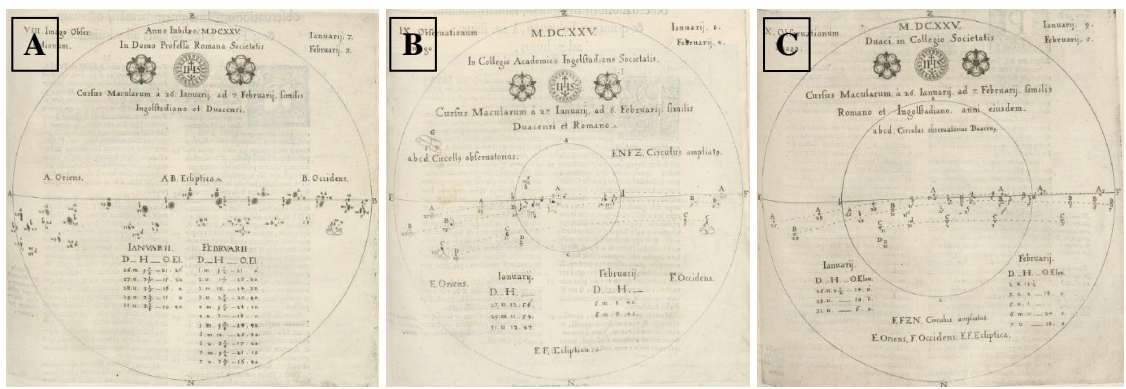

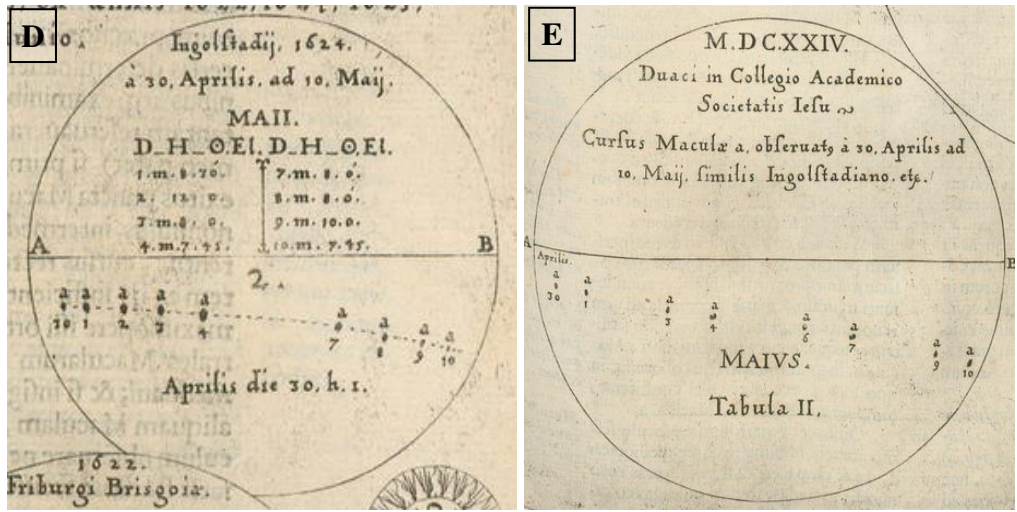

Figure 7. Sunspot drawings made by Scheiner (A), Schönberger (B), and Malapert (C) for the same observation period from 26 January to 6 February 1625 [Source: Scheiner, 1630, pp. 177, 179, and 181]. Sunspot group observed by Schönberger (D) and Malapert (E) from 30 April to 10 May 1624 [Source: Scheiner, 1630, pp. 217 and 229].

The sunspot drawing made by Malapert (1633) in March 1618 (Figure 5) can also be compared with those made by Cysat at Ingolstadt and Perovius at Kalisz (Figure 8). Perovius only observed on 9 March, recording 2 sunspots in one group, as also did Cysat, but Malapert did not record any observation. Instead, Malapert observed on two days, 8 and 10 March, when Cysat also did. According to the drawings, Cysat and Malapert recorded the same number of individual sunspots on 8 March (two sunspots in one group), but Malapert recorded a greater number of sunspots (four sunspots in one group) than Cysat (two sunspots in one group) on 10 March. This fact was noted in a comment by Malapert (1633, p. 70): [Original text] "Neuter autem in tam exiguo Solis disco minutiores huius cumuli particulas potuit perspicere." [English translation] "Neither observation [by Cysat and by Perovius] was capable of appreciating the tinier particles of this cumulus on such a small solar disk." Hence the sunspot observations made by Malapert seem more precise than at least those made by Cysat. We would also note that we do not know exactly whether, in this case, Malapert represented all the groups observed by Cysat and Perovius or only what was equivalent to group *A* in Malapert (1620) or *a* in Scheiner (1630). Thus, Malapert (1633, p. 70) reported: [Original text] "Vides observationem superioris maculae nostrae minore forma habitam Ingolstadii in Bavaria, et Callisii in Polonia." [English translation] "One can see the observation of our previous spot [Malapert, 1633, p. 69], at a smaller size, taken in Ingolstadt in Bavaria, and in Kalisz in Poland." It seems probable therefore that Malapert (1633) only showed one group on the sunspot drawings including records by Cysat and Perovius in order to compare with the only group that he recorded on his sunspot drawing.

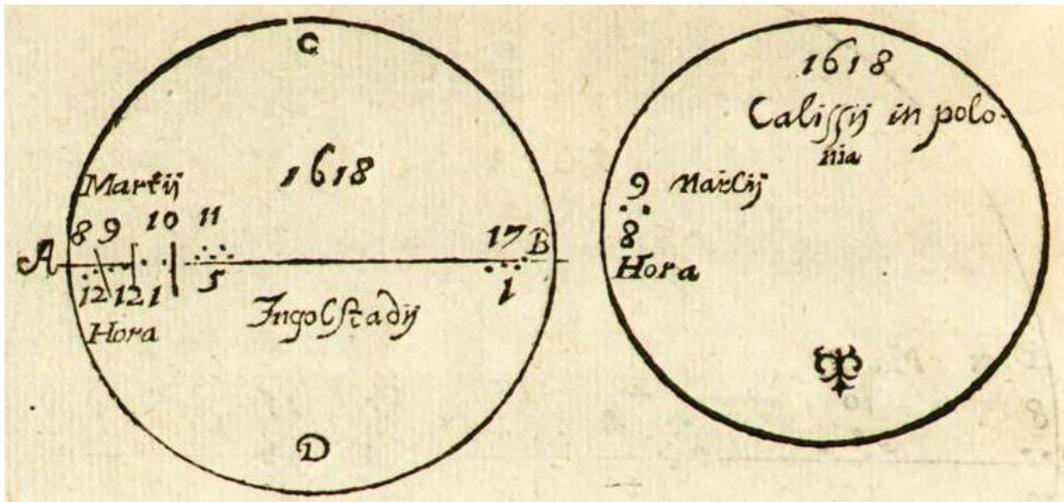

Figure 8. Sunspot drawings recorded by Cysat, at Ingolstadt (left), and Perovius at Kalisz (right) in March 1618 [Source: Malapert, 1633, p. 70].

We found another sunspot drawing with observations made by Perovius in July 1618 included in Malapert (1633, p. 75). In this case, there are two days (13 and 18 July) on which both Malapert and Perovius made sunspot observations. One can see that these two sets of observations are similar. On 13 July, Malapert and Perovius recorded three sunspots in one group, and, on 18 July, both astronomers recorded two sunspots in one group. The shape of the sunspots, however, seems slightly different in the two drawings. Malapert (1633) also reported sunspot observations made by Wely at Coimbra in October 1620 (Malapert, 1633, p. 78), December 1620 (Malapert, 1633, p. 80), and September 1621 (Malapert, 1633, p. 77). However, Malapert did not give sunspot drawings for Wely's observations, only textual reports. The comparison made by Malapert (1633) with respect to Wely's sunspot observations is with respect to the positions of the sunspots on the solar disk. Malapert (1633) reported that the sunspot observations made by Wely agree with his own records except on 7 December 1620, when Malapert (1633) indicated that the two observations differed significantly unless there had been an error in the values registered for the positions.

**5.5. Butterfly diagram**

We also constructed a butterfly diagram on the basis of the sunspot drawings made by Malapert (Figure 9, orange dots) published in Malapert (1633) and Scheiner (1630). Thus, we represent the heliographic latitude of all the sunspots recorded by Malapert versus the dates when they were observed. The positions of sunspots recorded by Malapert included in Scheiner (1630) were published by Arlt et al. (2016), and those obtained from Malapert (1633) were presented in Muñoz-Jaramillo & Vaquero (2018). We also depicted all the sunspot positions recorded by other observers as calculated by Arlt et al. (2016) from Scheiner (1630) (Figure 9, grey dots). The appearance of sunspots at high latitudes (35° approximately) in 1621 could indicate that a new solar cycle could well have started at around that year. We note that the minimum of the annual average of the group number obtained from Malapert's record also corresponds to 1621, although the temporal coverage for that year is low (Figure 4, bottom). In this way, these observations are framed within the first two solar cycles of the telescopic era. One cycle would include the first telescopic sunspot observations until 1621, and the second would start in 1621. Furthermore, for the period 1618–1620, one can see that most sunspot positions were observed in the northern hemisphere at heliographic latitudes between 20° and 10°. Sunspots appeared in the southern hemisphere at high latitudes at the end of 1620 and in 1621. In the final part of Malapert's records, from 1624 to 1626, sunspots were again recorded at lower latitudes (from 20° to 25°), and Malapert recorded more sunspots in the southern than in the northern hemisphere. In the last part of the second cycle, more sunspots again appeared in the northern hemisphere.

If one defines the normalized asymmetry as $NA = (G_N - G_S) / (G_N + G_S)$, where G is the group number and the subscripts N and S refer to the northern and southern hemispheres, one obtains an average NA equal to +0.27, corresponding to the period 1618 – 1621, and 0.05 for the second solar cycle of the telescopic era (1622 – 1631). We note that the NA for the period 1622-1627 is 0.32, clearly showing the southern hemisphere as being dominant in this period. All the sunspot positions shown in Figure 9 were used for the NA calculation. One notes that a strong hemispheric asymmetry occurred during the Maunder Minimum, a period characterized by a prolonged period of low solar activity from 1645 to 1715 (Eddy, 1976; Usoskin et al., 2015), since sunspots were mainly observed in the southern hemisphere (Ribes & Nesme-Ribes, 1993; Vaquero et al., 2015). However, the sunspot observations made by Hevelius just before the Maunder Minimum (1642-1645) do not show this asymmetry (Muñoz-Jaramillo & Vaquero, 2018).

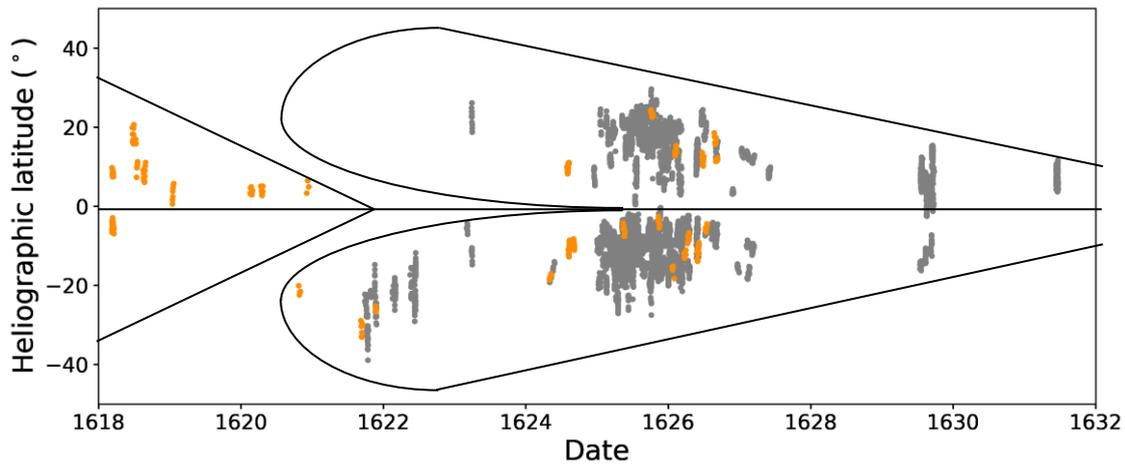

Figure 9. Butterfly diagram according to sunspot positions recorded by Malapert in the sunspot drawings analysed in this work (orange dots) and Scheiner (grey dots) analysed by Arlt et al. (2016). Black lines delimit the possible solar cycles during the period 1618-1631, and the northern and southern solar hemisphere.

## 6. Conclusions

We have carried out a revision of the sunspot records made by Charles Malapert during the period 1618–1626. The information studied in this work was published as sunspot drawings and textual reports in Malapert (1620, 1633) and Scheiner (1630). We analysed those documentary sources after translating the original Latin texts in order to provide an accurate analysis. Most of Malapert's sunspot observations are included in Malapert (1633) except for records corresponding to 1624 and the first third of 1625 which are in Scheiner (1630).

The method employed by Malapert for his observations consisted in projecting the solar disk onto a paper from a device that he constructed. Malapert was one of the most important sunspot observers for the period 1618–1626 in terms of the number of records (Vaquero et al., 2016). Although the temporal coverage of Malapert's records is not great, his sunspot observations are important because he was the only observer for some 60% of his observation days. We recovered 88 new records not included in the current group database, and corrected various errors corresponding, for example, to wrong dates and the daily number of groups. We would highlight that, as shown in Figure 4, the shape of a standard 11-year solar cycle can be intuited in this new accounting, unlike the case for the sunspot records included in the current group database. We would note that Malapert generally only recorded active days, except for some spotless days in 1624 (from 27 to 30 August, and 7 September). Furthermore, the average solar activity level obtained in this work from Malapert's records for the whole 1618–1626 period (1.44) is almost a third greater than that obtained from the current group database (1.09). We compared the sunspot records made by Malapert with observations made by other observers of that time – Scheiner in Rome, Cysat and Schönberger in Ingolstadt, Perovius in Kalisz, and Wely in Coimbra. In both their textual reports and sunspot drawings, one could see that Malapert's records were similar to those of the other observers in number of recorded sunspot groups and in sunspot positions. However, we would note that Scheiner's observations seem more precise since, for example, Malapert did not observe some groups close to the solar limb that were recorded by Scheiner. We also presented the butterfly diagram of the heliographic latitudes of sunspots recorded by Malapert. The beginning of a new solar cycle in 1621 seems to be intuitable since sunspots appeared at high latitudes in that year. The southern hemisphere was dominant from the beginning of the second solar cycle of the telescopic era until almost the end of that cycle when the northern hemisphere became dominant. The northern hemisphere was also dominant before 1622, at the end of the first solar cycle of the telescopic era.

Malapert (1633) always represented just a single sunspot group in his sunspot drawings, even though he sometimes observed several groups. An example of this fact can be found in the sunspots observed by Malapert in January 1626 (one group recorded) when he affirms that his sunspot observations agree exactly with sunspot records made by Scheiner who recorded between 3 and 5 groups for the same observation days. Therefore, we concluded that the sunspot accounting obtained from Malapert's records would represent the lower limit of the solar activity level. Thus, these records should be used with caution in characterizing the solar activity level of that time. This new accounting of the group number recorded by Malapert and the original Latin texts together with their English translations can be found on the HASO website: http://haso.unex.es.

**Acknowledgements**

This research was supported by the Economy and Infrastructure Counselling of the Junta of Extremadura through project IB16127 and grant GR18097 (co-financed by the European Regional Development Fund) and by the Ministerio de Economía y Competitividad of the Spanish Government (CGL2017-87917-P).

**Disclosure of Potential Conflicts of Interest** The authors declare that they have no conflicts of interest.